\documentclass{article}

\usepackage{arxiv}

\usepackage[utf8]{inputenc} 
\usepackage[T1]{fontenc}    
\usepackage{hyperref}       
\usepackage{url}            
\usepackage{booktabs}       
\usepackage{amsfonts}       
\usepackage{nicefrac}       
\usepackage{microtype}      
\usepackage{lipsum}		
\usepackage{graphicx}
\usepackage{doi}
\usepackage{enumitem}
\usepackage{makecell}

\title{Residual Fourier Neural Operator for thermochemical curing  modelling of composites}

\author{ 
     \\\large{Gengxiang Chen$^a$, Yingguang Li$^{a}$\thanks{Corresponding author. Email: liyingguang@nuaa.edu.cn,   Code available: \href{https://github.com/gengxiangc/ResFNO}{https://github.com/gengxiangc/ResFNO}},  Xu liu$^b$, Qinglu Meng$^a$, Jing Zhou$^a$, Xiaozhong Hao$^a$}
     \\
     \\ \small{$^a$ College of Mechanical and Electrical Engineering, Nanjing University of Aeronautics and Astronautics, Nanjing, China}
     \\ \small{$^b$ School of Mechanical and Power Engineering, Nanjing Tech University, Nanjing, China}

}

\renewcommand{\shorttitle}{}

\hypersetup{
pdftitle={A Residual Fourier Neural Operator for thermochemical curing  modelling of composite},
}

\begin{document}

\maketitle

\begin{abstract}

During the curing process of composites, the temperature history heavily determines the evolutions of the field of degree of cure as well as the residual stress, which will further influence the mechanical properties of composite, thus it is important to simulate the real temperature history to optimize the curing process of composites. Since thermochemical analysis using Finite Element (FE) simulations requires heavy computational loads and data-driven approaches suffer from the complexity of high-dimensional mapping. This paper proposes a Residual Fourier Neural Operator (ResFNO) to establish the direct high-dimensional mapping from any given cure cycle to the corresponding temperature histories. By integrating domain knowledge into a time-resolution independent parameterized neural network, the mapping between cure cycles to temperature histories can be learned using limited number of labelled data. Besides, a novel Fourier residual mapping is designed based on mode decomposition to accelerate the training and boost the performance significantly. Several cases are carried out to evaluate the superior performance and generalizability of the proposed method comprehensively. 

\end{abstract}


\section{Introduction}
Composite materials, which consist of fibers (e.g. carbon fiber and glass fiber)
embedded in polymer matrices (e.g. epoxy resin and unsaturated polyester resin),
can provide higher structural performance (e.g. strength-to-weight ratio and
corrosion resistance) than traditional metallic materials, and have been widely
used in many fields including aerospace, automotive and renewable energy \cite{harris2002design} \cite{crawford2021mini}.
To manufacture a composite structure, fibers impregnated with partially cured
matrices are first cut and stacked to form a component of desired shape. This
preformed component then have to be cured during which the matrices undergo the
polymerization reaction initiated by heating following an appropriate cure
cycle, to achieve the required mechanical properties \cite{chen2021effect}. The cure cycle
heavily determines the evolutions of the fields of temperature and degree of
cure within the composite component which will further influence the residual
stress field, thus it has to be seriously designed otherwise detects like
incomplete cure, severe temperature overshoots, unacceptable levels of residual
stresses and distortion may happen which will greatly degrade the performance of
the final composite structure \cite{struzziero2019numerical}\cite{hubert2012autoclave}.

The model that maps cure cycles to the fields of temperature and degree of cure
is essential for designing optimal cure cycles \cite{chen2021effect}. Though the physics of the
thermochemical curing process is well established and modeled by a set of
coupled nonlinear partial differential equations (PDEs) describing heat
conduction and resin cure kinetics, the closed-form solution is unavailable \cite{hubert2012autoclave}.
Therefore, the fields of temperature and degree of cure nowadays are usually
calculated by computational approximation like finite element method (FEM) \cite{carrera2002theories}.
But for parts with large size and complex structures, the computational
efficiency of FEM is unable to meet the requirement for the iterative
optimization of cure cycles in practical applications \cite{fernlund2003finite}.

Recently, data driven models like neural network are researched to approximate
the mapping by training with the data samples generated from expensive
high-fidelity simulations with FEM \cite{zhang2021physical}\cite{chen2019pose}. These attempts are meaningful because
an accurate data-driven surrogate model can greatly speed-up the optimization of
curing process to minimize the risk in manufacturing \cite{humfeld2021machine}. Zobeiry et al. \cite{zobeiry2021physics}
established several data-driven models based on FE simulations to predict curing
related parameters including steady-state thermal lag, transient thermal lag and
exotherm. In their pioneering work, comprehensive domain knowledge was
integrated into the components of neural network models to optimize feature
transformation and activation function. To further reduce the data requirements,
Ramezankhani et al. \cite{ramezankhani2021making} proposed an interesting transfer learning framework to
transfer the learned exotherm prediction model from one-hold cure cycle
condition to two-hold cure cycle condition with only 500 simulated data for the
target domain. Furthermore, they studied an active learning strategy to guide
FEM simulation process so that the exotherm and thermal lag prediction models
can be trained using only limited amount of data \cite{ramezankhani2021active}.

The researches mentioned above reveal that data-driven models are capable of
achieving high accuracy result with small datasets in composite thermal
analysis. However, these works focus on the prediction of some thermal
parameters rather than the real temperature history of composites. Since the
cure cycles and temperature histories are both functions of time and
conventional machine learning methods can only build mapping from limit number
of input features to few output features, it is difficult or impossible to train
a high-dimensional mapping from the cure cycles to temperature or degree of cure
histories directly. Therefore, cure cycles are usually simplified using few
specific features including heat rate and top temperature, and the required
temperature history is usually replaced by some representative parameters such
as thermal lag and exotherm \cite{zobeiry2021theory}\cite{ramezankhani2021making}. Keith et al. \cite{humfeld2021machine} represented the
temperature histories and cure cycles as corresponding time series and designed
a Long Short-Term Memory (LSTM) model to predict the temperature for each moment
using time sequence evolution. However, the accuracy of this LSTM model relies
heavily on a large amounts of simulation data (more than 100,000) generated by
FEM software, which will restrict the application in real engineering scenarios.

Physics-informed Neural Networks (PINNs) emerge as an alternative approach to
solve complex engineering problems \cite{chen2021physics}. By designing an integrated loss
function based on PDE functions of the system, initial condition, and boundary
condition, PINN can learn the underlying model by self-training without labelled
data. Equations and conditions supervise the training instead of the traditional
prediction error from labelled data \cite{raissi2017physics}. Sina et al. \cite{niaki2021physics} presented an
elaborate PINN to simulate the fields of temperature and degree of cure of a
composite-tool system. The coordinates of time and space are input features and
the corresponding degree of cure and temperature can be predicted. Similarly,
Zobeiry et al. \cite{zobeiry2021physics}] developed a PINN model to solve the temperature over the
curing period and evaluated its effectiveness in several cases. Although these
methods can solve heat transfer PDEs accurately subjected to the specific cure
cycle, the model needs to be retrained completely for any new cure cycle, which
cannot satisfy the requirements of cure cycles designing and optimization.

As reviewed above, it is still a great challenge to build a model to predict the
entire temperature and the degree of cure histories for any given cure cycle
because of the complexity of high-dimensional mapping. Recently, neural
operators emerged as a new concept by generalizing standard feed-forward neural
networks to learn mappings between infinite-dimensional spaces of functions
without increasing the complexity of the network \cite{kovachki2021neural}. Based on the neural
operator theory, this paper proposes a Residual Fourier Neural Operator (ResFNO)
to establish the direct high-dimensional mapping from any given cure cycle to
the corresponding temperature histories. A novel Fourier residual mapping is
designed based on domain knowledge to accelerate the training and boost the
performance. Several cases are carried out to evaluate the accuracy and
generalizability of the proposed method comprehensively.

The reminder of the paper is organized as follows: The background of heat
transfer problem of composite is introduced in section 2. The main framework of
the proposed ResFNO is presented in detail in Section 3. Experimental results
and analysis of 3 cases are provided in Section 4. The main conclusions are
presented in the final section.

\section{Background}
The general form of exothermic heat transfer in composites curing process can be expressed as the following PDE:

\begin{equation}
\rho C \frac{\partial T}{\partial t}=\frac{\partial}{\partial x}\left(k_{x} \frac{\partial T}{\partial x}\right)+\frac{\partial}{\partial y}\left(k_{y} \frac{\partial T}{\partial y}\right)+\frac{\partial}{\partial z}\left(k_{z} \frac{\partial T}{\partial z}\right)+\dot{Q}
\end{equation}

Where $T$ is the temperature, $\rho$ and $C$ are density and specific heat capacity, and $k$ indicates the thermal conductivity in specific direction. $\dot{Q}$ is the internal heat source, i.e. the rate of heat generation caused by the exothermic polymerization reaction in composite part, which is determined by instantaneous temperature and degree of cure. Besides, the initial conditions and boundary conditions are necessary for solving the PDE mentioned above. The typical boundary conditions include: Dirichlet boundary, Neumann boundary and Robin boundary.

\begin{itemize}[leftmargin=*]

\item  Initial condition: 
\begin{equation}
T_{t=0}=T_{0} ; \alpha_{t=0}=\alpha_{0}
\end{equation}

\item Dirichlet boundary: 
\begin{equation}
T_{x_b}=T_{a}(t)
\end{equation}

\item Neumann boundary: 
\begin{equation}
\left.k_{x} \frac{\partial T}{\partial x}\right|_{x_b}=q_{x_b}
\end{equation}

\item Robin boundary: 
\begin{equation}
h\left(T_{x_b}-T_{a}(t)\right)=\left.k_{x} \frac{\partial T}{\partial x}\right|_{x_b}
\end{equation}

\end{itemize}

Where $T_{0}$ and $\alpha_{0}$ are the initial temperature and degree of cure and $q$ is the flux density on the boundary. Equivalent to the air temperature in the autoclave and surfaces temperature in RTM (Resin transfer Molding), $T_{a}(t)$ indicates the external temperature.

\begin{figure}[th]
	\centering
	\includegraphics[width=1\linewidth]{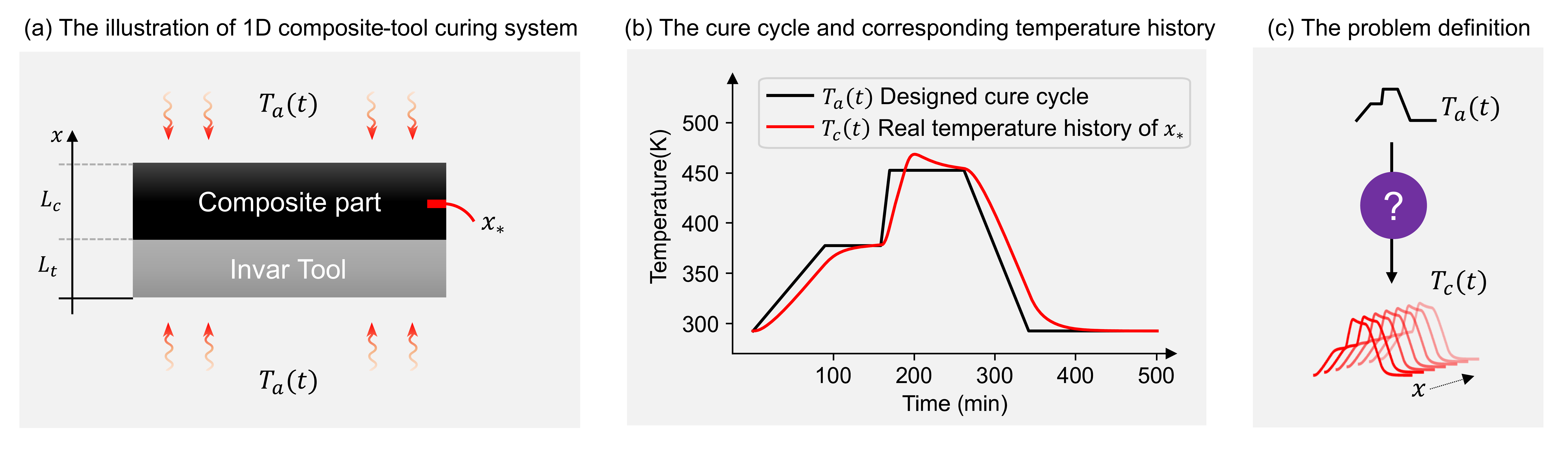}\\[-2mm]	
	\caption{Background of composites curing and problem definition}
	\label{fig:fig1}
\end{figure}

In this study, we start with an example of $1 \mathrm{D}$ composite-tool curing system with convective boundary conditions (Neumann boundary) in autoclave shown in Fig. \ref{fig:fig1}(a). The heat transfer governing equation and boundary conditions can be formulated as:

\begin{equation}
\rho C \frac{\partial T}{\partial t}=\frac{\partial}{\partial x}\left(k_{x} \frac{\partial T}{\partial x}\right)+\dot{Q} 
\end{equation}
\begin{equation}
h_{t}\left(T_{x=0}-T_{a}(t)\right)=\left.k_{t} \frac{\partial T}{\partial x}\right|_{x=0} \\
\end{equation}
\begin{equation}
h_{c}\left(T_{a}(t)-T_{x=L_{t}+L_{c}}\right)=\left.k_{c} \frac{\partial T}{\partial x}\right|_{x=L_{t}+L_{c}}
\end{equation}

Where $h_{c}\left(h_{t}\right)$ is the heat transfer coefficient of composite (tool) and $T_{x=0}\left(T_{x=L_{t}+L_{c}}\right)$ is the temperature of composite-tool system at its lower (upper) surface. To guide the curing process, a cure cycle of air, i.e. $T_{a}(t)$ depicted in Fig. \ref{fig:fig1}(a), is designed to ensure the end-part quality. During the heat-up and cool-down stages, composite part temperature $T_{c}(t)$ lags behind $T_{a}(t)$ due to the thermal masses and thermal resistances of the part and tool [1]. However, as exothermic curing reaction starts within the part, $T_{c}(t)$ gradually increases beyond $T_{a}(t)$ and reaches the maximum temperature, as shown in Fig. \ref{fig:fig1}(b). The temperature field directly determines the evolutions of the fields of degree of cure as well as the residual stress, which will further influence the mechanical properties of the cured composite, thus it is important to simulate the entire temperature history to optimize the curing process.

As shown in Fig. \ref{fig:fig1}(c), the purpose of this research is to build a surrogate model to predict the temperature histories $T_{c}(t)$ of all points on the composite-tool system for any given cure cycle function $T_{a}(t)$ accurately and effectively. Since the cure cycles and temperature histories are both functions of time, it is difficult or impossible to train a high-dimensional mapping from the cure cycles to temperature histories directly using conventional machine learning methods. Thus, the challenges of this research involve how to represent the input $T_{c}(t)$ and the output $T_{a}(t)$, how to integrate domain knowledge to accelerate the training, and how to learn an accurate prediction model using only limited simulation data.

\section{ResFNO framework}
\label{sec:others}
\begin{figure}[th]
	\centering
	\includegraphics[width=1\linewidth]{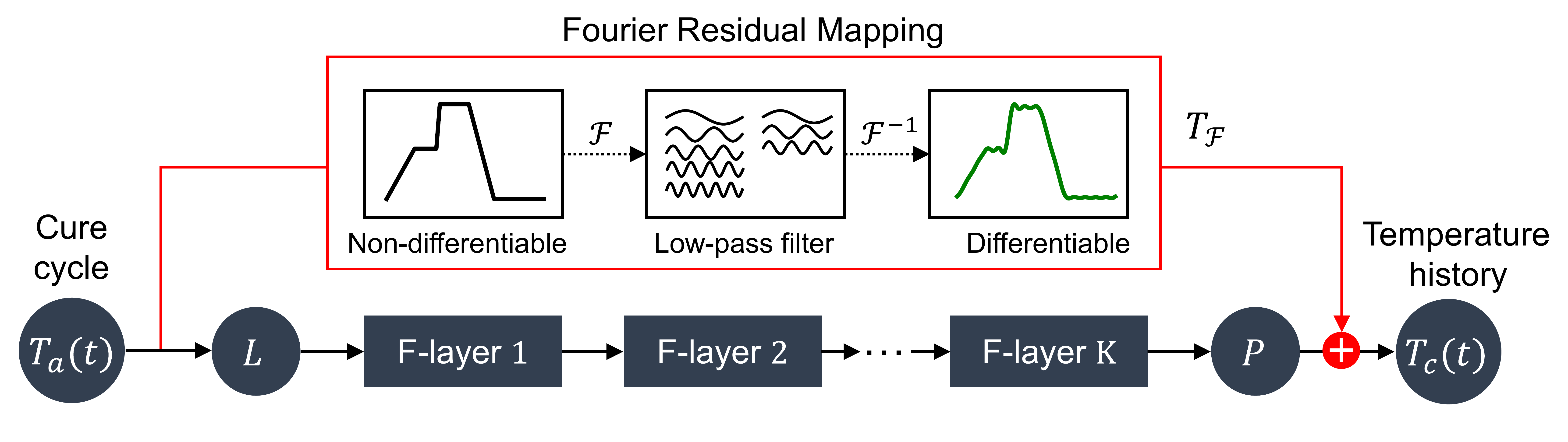}\\[-2mm]	
	\caption{Residual Fourier Neural Operator framework for temperature history prediction}
	\label{fig:fig2}
\end{figure}

As shown in Fig. \ref{fig:fig2}, a Residual Fourier Neural Operator (ResFNO) framework is proposed to establish a direct mapping from the cure cycle $T_{a}(t)$ to the temperature history $T_{c}(t)$. In section 3.1, we introduce the typical Fourier Neural Operator method involving neural operator theory and Fourier layers, and explain how and why FNO can potentially predict the temperature history. In section 3.2, a novel Fourier residual mapping is presented to improve the typical FNO using domain knowledge.

\subsection{Fourier Neural Operator}
\subsubsection{Neural Operators}

When the cure period is discretized into finite number of time intervals, the cure cycle $T_{a}$ can be expressed by a vector sampled from the vector space of cure cycles over $\mathbb{R}^{n_{t}}$, where $n_{t}$ is the size of time intervals. Therefore, existing researches predict the temperature of subsequent time with LSTM by modeling the cure cycle as a discrete time series. From a more general perspective, both the cure cycle $T_{a}(t)$ and the temperature history $T_{c}(t)$ are functions sampled from two unknown temperature function spaces, thus the mapping from $T_{a}(t)$ to $T_{c}(t)$ can be treated as an operator between infinite dimensional function spaces. If the operator can be learned using a finite collection of input-output pairs $\mathcal{D}=$ $\left\{\left(T_{a}(t)_{1}, T_{c}(t)_{1}\right), \ldots,\left(T_{a}(t)_{n}, T_{c}(t)_{n}\right)\right\}$, the temperature history $T_{c}(t)^{*}$ can be evaluated directly with a given cure cycle $T_{a}(t)^{*}$ rather than performing time-consuming simulation or series prediction. However, the mapping from $T_{a}(t)^{*}$ to $T_{c}(t)^{*}$ is a high-dimensional inputoutput task, which is a challenging machine learning problem because the number of parameters of the networks depends on the resolution of input data. High-resolution time domain data can greatly increase the complexity of the networks and the performance of the model will rely on huge amount of training data. Therefore, it is necessary to build a resolution-independent framework to learn high-dimensional mappings between functions.

Recently, neural operators emerged as a new concept by generalizing standard feedforward neural networks to learn mappings between infinite-dimensional spaces of functions [18]. Infinite-dimension means that the operators can process very high-dimensional mapping without increasing the complexity and the number of network parameters, which can bring a new solution for data-driven PDEs, including the heat transfer problems in composite curing. With the given collections of cure cycles, the learning process of neural operators could be regard as solving the empirical-risk minimization problem, which is equivalent as that of standard machining learning problem. The target function can be represented as:

\begin{equation}
\min _{\theta} \frac{1}{n} \sum_{i=1}^{n}\left\|T_{c}(t)_{i}-\mathcal{G}_{\theta}\left(T_{a}(t)_{i}\right)\right\|_{u}^{2}
\end{equation}

Where $G_{\theta}$ is the neural operator parameterized by $\theta$. The concrete representation of $G_{\theta}$ is normally a series of linear operators and non-linear operators, including lifting layer, iterative kernel integration layer and projection layer.

\begin{figure}[th]
	\centering
	\includegraphics[width=1\linewidth]{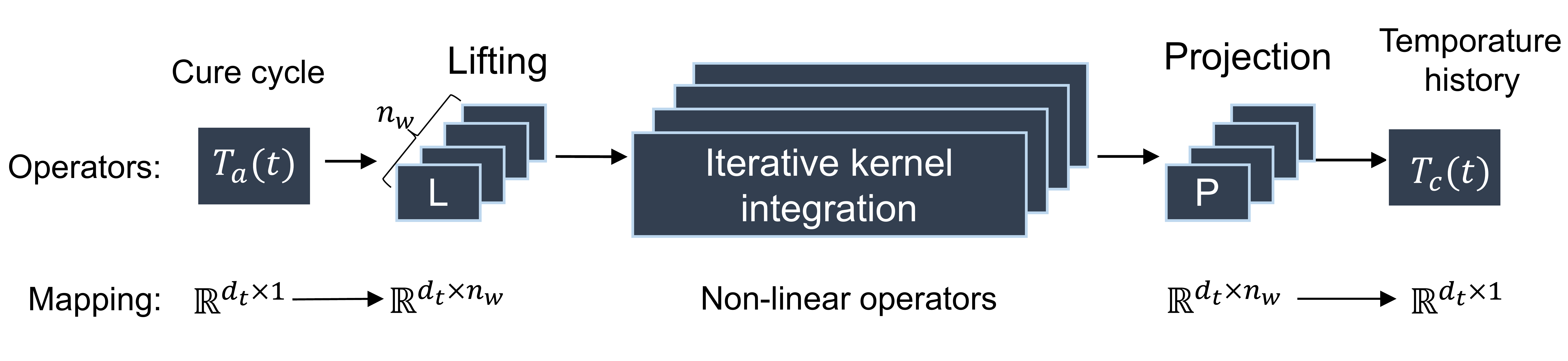}\\[-2mm]	
	\caption{Typical neural operator framework }
	\label{fig:fig3}
\end{figure}

As shown in Fig. \ref{fig:fig3}, the input cure cycle is $T_{a}(t) \in \mathbb{R}^{d_{t} \times 1}$ where $d_{t}$ depends on the resolution of time intervals. This input structure seems consistent with traditional deep learning method. The great difference is that the dimension $d_{t}$ never participate the transformation operation in linear operators and non-linear operators. The lifting operator $(L)$ and projection operator $(P)$ are parameterized by two simple linear transformation $\theta_{L} \in \mathbb{R}^{1 \times n_{w}}$ and $\theta_{P} \in$ $\mathbb{R}^{n_{w} \times 1}$, where $n_{w}$ is the width of the neural operator. The lifting operator aims to add more channels to enhance the representation capability of the neural network. In the next section, it will be further explained that the non-linear operators are also parameterized independent with $d_{t}$. Therefore, the complexity and size of parameters of the neural operator will not be influenced by the size of $d_{t}$ despite $d_{t}$ can be very large. This characteristic allows us to train the neural operator with low time-resolution dataset, and then predict on high time-resolution cure cycles. The neural operator from $T_{a}(t)$ to $T_{c}(t)$ can be treated as approximating the potential operator $\mathcal{G}_{\theta}$ by linear combination of limited number of non-linear sub-operators:

\begin{equation}
g_{\theta}\left(T_{a}(t)\right)=\sum_{j=1}^{n_{w}} \theta_{L j} \varphi_{j}(t) \theta_{P j}
\end{equation}

Where $\varphi_{j}(t)$ is the $j$-th iterative kernel integration, $\theta_{L j}$ and $\theta_{P j}$ are parameters of $j$-th channel in the lifting layer and projection layer.

\subsubsection{Fourier Layer}

Iterative kernel integration using Fourier layers are proven to be expressive enough to approximate any measurable operator mapping. The iterative updates process from Fourier layer- $i$ to Fourier layer $i+1$ can be simply represented as:

\begin{equation}
v_{i+1}(t):=\sigma\left(W v_{t}(x)+R \cdot \mathcal{F}_{*} v_{i}(w)\right)
\end{equation}

\begin{figure}[th]
	\centering
	\includegraphics[width=1\linewidth]{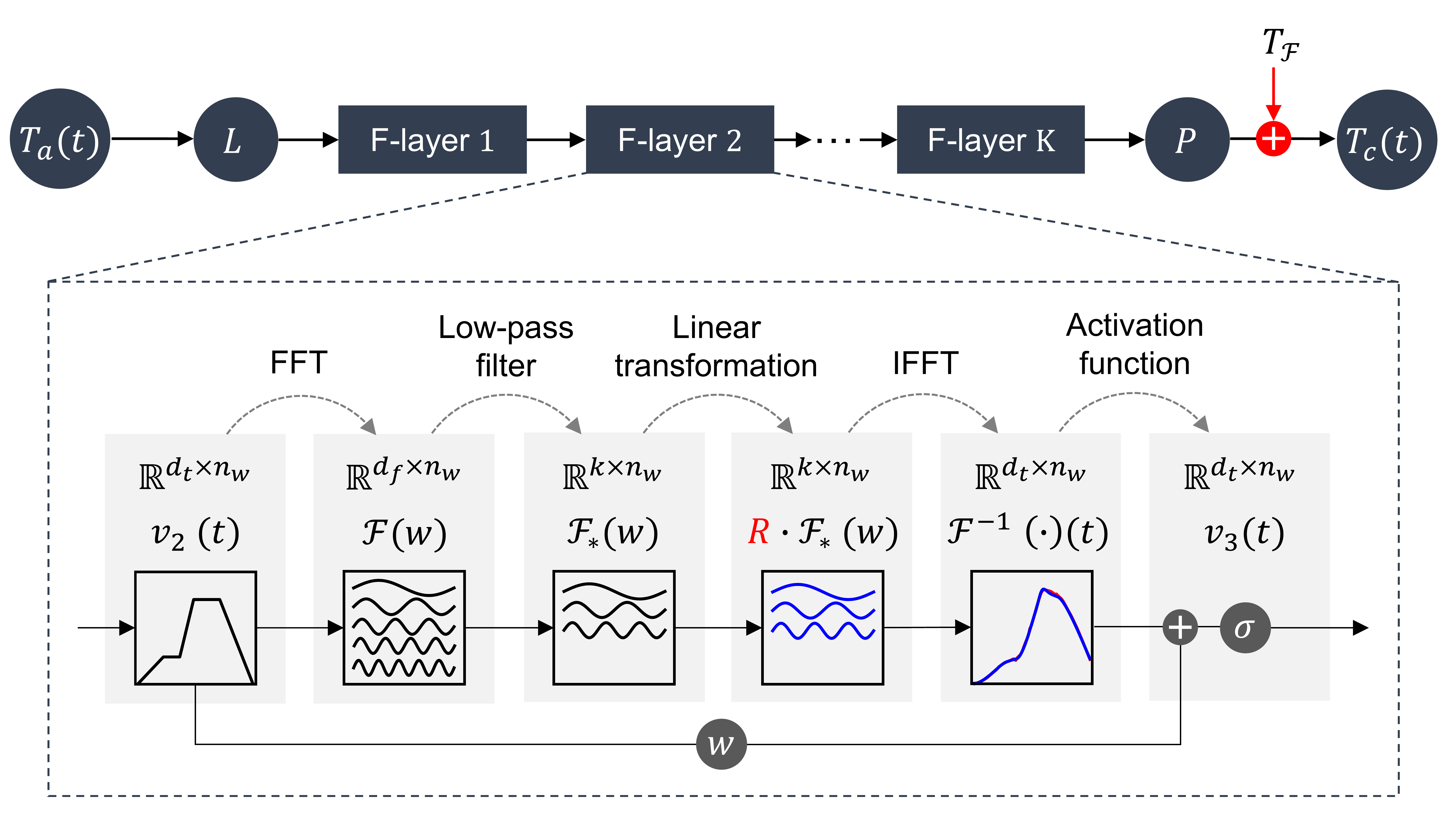}\\[-2mm]	
	\caption{Fourier layer }
	\label{fig:fig4}
\end{figure}

The detailed derivations of Fourier layer can be found in $[2]$, which is not in the slope of this research. Here, we will introduce how and why Fourier layer is efficient to establish the mapping between the input cure cycle and the output temperature history. As shown in Fig. \ref{fig:fig4} , a standard Fourier layer consists of five steps, FFT (Fast Fourier Transform), Low-pass filter, Linear transformation, IFFT (Inverse Fast Fourier Transform), and non-linear activation.

\paragraph{Step1: FFT} Denote $v(t) \in \mathbb{R}^{d_{t}} \times n_{w}$ as the features after the lifting layer, $v(t)$ can be treated as $n_{w}$ time-domain functions $\left\{v^{1}(t), \ldots, v^{n_{w}}(t)\right\}$ because $d_{t}$ is the discrete size of the curing time. After FFT, time-domain functions for all channels are represented as $n_{w}$ frequency domain functions $\left\{\mathcal{F}^{1}(t), \ldots, \mathcal{F}^{n_{w}}(t)\right\}$. The frequency modes of the cure cycle functions can be extracted as new features of the neural network which can be more representative than purely time domain features. Note that, the number of parameters of the subsequent layers will also increase if all the frequency features are directly propagated to the next layer because the features after FFT is $\mathcal{F}(t) \in \mathbb{R}^{d_{f} \times n_{w}}$, where $d_{f}=1+d_{t} / 2$.

\paragraph{Step2: Low-pass filter} Normally, the cure cycles are characterized by heating stages, cooling stages and hold stages, and all these characteristics can be treated as low frequency features in the frequency domain. Therefore, it is reasonable to keep low frequency modes of $v(t)$ and abandon high frequency modes. Suppose $k$ lower modes are selected to propagate to next layer, the feature after low-pass filter is $\mathcal{F}_{*}(w) \in \mathbb{R}^{k \times n_{w}}$.

\paragraph{Step3: Linear transformation} A linear transformation is parameterized for each channel as $R^{j} \in \mathbb{R}^{k \times k}$ for $j=1,2, \ldots, n_{w}$ so that the $k$ input Fourier modes can be transferred to another appropriate $k$ Fourier modes. The total linear transformation layer can be parameterized as a tensor $R \in \mathbb{R}^{k \times k \times n_{w}}$, which is only dependent on the number of truncated Fourier modes $k$ and the number of channels $n_{w}$. The linear transformation in Fourier space can capture high-nonlinear features in the original space with resolution-invariant parameters.

\paragraph{Step4: IFFT} Fourier features will be transformed to time domains by non-parametric IFFT operator. Then the feature becomes $\mathcal{F}^{-1}(\cdot)(t) \in \mathbb{R}^{k \times n_{w}}$ and the dimension returns to $k \times n_{w}$

\paragraph{Step5: Non-linear activation} This step consists of a local linear transform $W$ and an activation function, which play the same roles as in traditional full-connected neural networks. The activation function can be sigmoid, Relu, or tanh.

After the five-steps Fourier layer, the input feature $v_{i}(w)$ become $v_{i+1}(w)$. Normally, the Fourier layer will be repeated several times to strengthen the representative performance.

\subsection{Fourier Residual Mapping}
Typical FNO can be used to build high-dimensional mapping for different kinds of engineer problems as long as enough labeled data is available. However, it is expensive and time-consuming to collect the datasets of temperature histories both experimentally and computationally. In this section, a novel Fourier residual mapping is presented by combining the data-drive model and domain knowledge, thus the new framework can build the prediction model with high performance using only few data.

The first inspiration comes from the fact that the real temperature history is always highly correlated with the designed cure cycle. Although there are thermal lag and exotherm during heating, the basic shape and tendency over times are consistent. As analyzed in ResNet, it is easier to learn a residual function with reference to the input than learn an unreferenced function \cite{he2016deep}. Denoting the underlying mapping from $T_{a}(t)$ to the temperature history $T_{c}(t)$ as $\mathcal{G}_{\theta}\left(T_{a}(t)\right)$, let the network learn another mapping $\mathcal{H}_{\theta}\left(T_{a}(t)\right)=\mathcal{G}_{\theta}\left(T_{a}(t)\right)-T_{a}(t)$. From the perspective of function space, the original mapping $G_{\theta}\left(T_{a}(t)\right)$ tends to be unrestricted and there are infinite number of potential solutions distributed over the whole function spaces. However, the mapping $\mathcal{H}_{\theta}\left(T_{a}(t)\right)$ is easier to learn because it is closer to zero mapping. As shown in Fig. \ref{fig:fig2}, the residual mapping can be realized by FNO networks with a 'shortcut connections' from the input to the output.

\begin{figure}[th]
	\centering
	\includegraphics[width=1\linewidth]{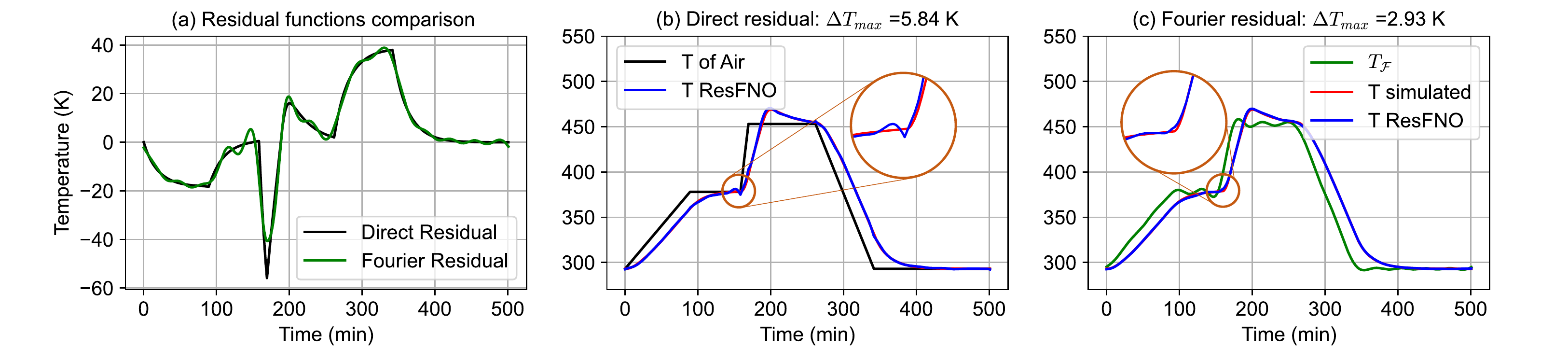}\\[-2mm]	
	\caption{Comparison between original residual and Fourier residual}
	\label{fig:fig5}
\end{figure}

If we go deeper into the residual term $r(t)=T_{c}(t)-T_{a}(t)$, one great challenge is that $r(t)$ is non-differentiable because the cure cycle is a non-differentiable piecewise function. The residual term $r(t)$ is represented by the black curve in Fig. \ref{fig:fig5}(a) and the corresponding $T_{c}(t)$ and $T_{a}(t)$ are red curve and black curve in Fig. \ref{fig:fig5}(b) respectively. It is obvious that there are 5 non-differentiable points on $T_{a}(t)$ and $r(t)$.

The output of FNO is differentiable because it is reconstructed by limited number of frequency modes. Therefore, there will be inevitable approximate error if we use a differentiable function to fit a non-differentiable function $r(t)$. The predicted temperature history using direct residual FNO is shown as the blue curve in Fig. \ref{fig:fig5}(b). There is a big prediction error $(5.84 \mathrm{~K})$ at the turning points of $T_{a}(t)$ (around $\left.t=160 \mathrm{~min}\right)$.

Inspired by the Fourier layer, a novel Fourier residual mapping is proposed to tackle the abovementioned challenge. As shown in Fig. \ref{fig:fig2}, the original cure cycle $T_{a}(t)$ can be transferred to a differential function $T_{\mathcal{F}}$ by reconstructing the signal using lower frequency modes. The number of modes can be the same with that in Low-pass filter in Fourier layer, namely $k$. Then the Fourier residual can be represented as $r_{\mathcal{F}}(t)=T_{c}(t)-T_{\mathcal{F}}$. The reconstructed cure cycle $T_{\mathcal{F}}$ and Fourier residual $r_{\mathcal{F}}(t)$ are shown as the green curve in Fig. \ref{fig:fig5}(c) and Fig. \ref{fig:fig5}(a) respectively. It is obvious that both functions become differentiable. The temperature history predicted using Fourier residual is shown as the blue curve in Fig. \ref{fig:fig5}(c). It can be observed that the singular error in the non-differentiable point is greatly reduced.

The proposed framework ResFNO is simple but effective. It will be demonstrated in the following experiments that the integration of domain knowledge can reduce the testing error, accelerate the training process as well as reduce the requirements of training data.

\section{Implementation and validation}
\paragraph{Configuration of composites}  To evaluate the performance of the proposed method, we consider a $1 \mathrm{D}$ exothermic heat transfer situation where an AS4/8552 composite part with a thickness of $L_{c}=20 \mathrm{~mm}$, is placed on an Invar tool with a thickness of $L_{t}=30 \mathrm{~mm}$. The material properties for AS4 fiber, 8552 epoxy and Invar tool are listed in Table 1. Given the densities and specific heat capacity of fiber and resin, the density $\rho_{c}$ and specific heat capacity $C_{c}$ of composite can be calculated via rules of the mixtures, i.e., Eq. (\ref{rho}) and Eq. (\ref{cccv}). The thermal conductivity of the composite part in thickness direction can be obtained from Springer-Tsai model \cite{springer1967thermal}, as shown in Eq. (\ref{kc}), Eq. (\ref{gamma}) and Eq. (\ref{omega}). Considering the complex flow field inside autoclaves, we assume convective boundary conditions with different heat transfer coefficients above composite part $h_{c}=$ $120 \mathrm{~W} / \mathrm{m}^{2} \mathrm{~K}$ and under the tool $h_{t}=70 \mathrm{~W} / \mathrm{m}^{2} \mathrm{~K}$.

\begin{table} \
	\caption{Material properties for AS4 fiber, 8552 epoxy and Invar tool}
	\centering
	\vspace{-0.5em}
    \begin{tabular}{lcccc}
     \toprule & Volume & Density, & Specific heat capacity, & Thermal conductivity, \\
    & fraction, $v$ & $\rho\left(\mathrm{kg} / \mathrm{m}^{3}\right)$ & $C(\mathrm{~J} / \mathrm{kg} \mathrm{K})$ & $k(\mathrm{~W} / \mathrm{m} \mathrm{K})$ \\
     \midrule
    AS4 fiber & $v_{f}=0.574$ & $\rho_{f}=1790$ & $C_{f}=914.0$ & $k_{f}=3.960$ \\
    8552 resin & $v_{r}=0.426$ & $\rho_{r}=1300$ & $C_{r}=1304.2$ & $k_{r}=0.212$ \\
    Invar tool & $-$ & $\rho_{t}=8150$ & $C_{t}=510.0$ & $k_{t}=13.0$ \\
    \bottomrule
    \end{tabular}
    \vspace{-0.4em}
	\label{tab:table1}
\end{table}

\begin{equation} \label{rho}
\rho_{c}=\rho_{r} v_{r}+\rho_{f} v_{f}
\end{equation}

\begin{equation} \label{cccv}
C_{c}=C_{r} v_{r}+C_{f} v_{f}
\end{equation}

\begin{equation} \label{kc}
k_{c}=k_{r}\left((1-2 \Omega)+\frac{1}{\Gamma}\left(\pi-\frac{4}{r} \tan ^{-1}\left(\frac{\sqrt{1-\Gamma^{2} \Omega^{2}}}{1+\Omega \Gamma}\right)\right)\right)
\end{equation}

\begin{equation} \label{gamma}
\Gamma=2\left(\frac{k_{r}}{k_{f}}-1\right)
\end{equation}

\begin{equation} \label{omega}
\Omega=\sqrt{\frac{v_{f}}{\pi}}
\end{equation}

\paragraph{Cure cyles of 3 cases}  Three group of cure cycles are designed for the composite-tool system: extreme two-hold temperature cycle, realistic two-hold temperature cycle and smart cure temperature cycle \cite{soohyun2015smart}. The principles of designing cure cycles generally include heating rate, holding temperature, and holding time, which are relevant to heating methods and the resin type. Extreme two-hold temperature cycles generated by exaggerated process parameters doesn't consider practical application, which is used to check the performance of the proposed method. As for realistic two-hold temperature cycles, we take the common heating rate range of autoclave, post-cure temperature and holding time into consideration. Smart cure cycles composed of cure-triggering, cooling and post-cure process are developed to reduce the thermal residual stress in some specific scenarios \cite{soohyun2015smart}. In this study, we set different variable ranges based on the temperature curve suggested by prepreg manufacturer. Moreover, we also replace the heating/cooling rate parameters with 'heating time' to simplify the generation of curves. Detailed variable ranges of three temperature cycles are listed in Table 2. All curves are generated by randomly selecting variables from predefined ranges. The results of schematic temperature cycles are depicted in Figure 1 . The corresponding thermal histories of tool and composite are simulated in Comsol $5.4$ software.

\begin{table}
	\caption{Detailed variable ranges of three temperature cycle cases }
	\centering
	\vspace{-0.5em}
    \begin{tabular}{c c l} 
    \toprule        &  $\Delta t(\min )$      & $\Delta T(K)$ \\
     \midrule     Case 1 &
    \makecell[c]{$t_{1} \in(20,80); t_{2} \in\left(t_{1}+20,110\right) ;$ \\ $t_{3} \in\left(t_{2}+20,150\right)$;\\ $t_{4} \in\left(t_{3}+20,190\right);$ \\$t_{5}=222;$}&    \makecell[c]{$T_{0}=293$ \\ $T_{1} \in(50,120)$\\ $T_{2} \in(100,240)$} \\
    
     \midrule   Case 2 & 
      \makecell[c]{$t_{1} \in(40,90)$ ; $t_{2} \in\left(t_{1}+30, t_{1}+90\right)$ \\ 
    $t_{3} \in\left(t_{2}+10, t_{2}+40\right) ;$  $T_{1} \in(80,140)$  \\$t_{4} \in\left(t_{3}+90, t_{3}+150\right) ;$ \\ $t_{5}=t_{4} + T_{2}/2$; $t_{6} = 500$}
    & \makecell[c]{$T_{0}=293$\\ $T_{1} \in(80,140)$\\ $T_{2} \in(150,170)$|} \\ 
    
     \midrule    Case 3 & 
    \makecell[c]{$t_1\in\left(30,120\right)$; $t_2\in(t_1+30,t_1+80);$
    \\$t_3\in(t_2+10, t_2+30)$; \\$t_4\in(t_3+10,t_3+20)$;\\$t_5\in(t_4+30,t_4+60)$;
    \\$t_6\in(t_5+90,t_5+150)$;\\$t_7\in(t_6+20,t_6+60)$; \\$t_8=450$;}
    & \makecell[c]{$T_0=293$ \\ $T_1\in(140,190)$ \\ $T_2\in(30,50)$ \\$T_3\in(80,140)$} \\
    \bottomrule
    \end{tabular}
	\label{tab:table2}
\end{table}

\begin{figure}[th]
	\centering
	\includegraphics[width=1\linewidth]{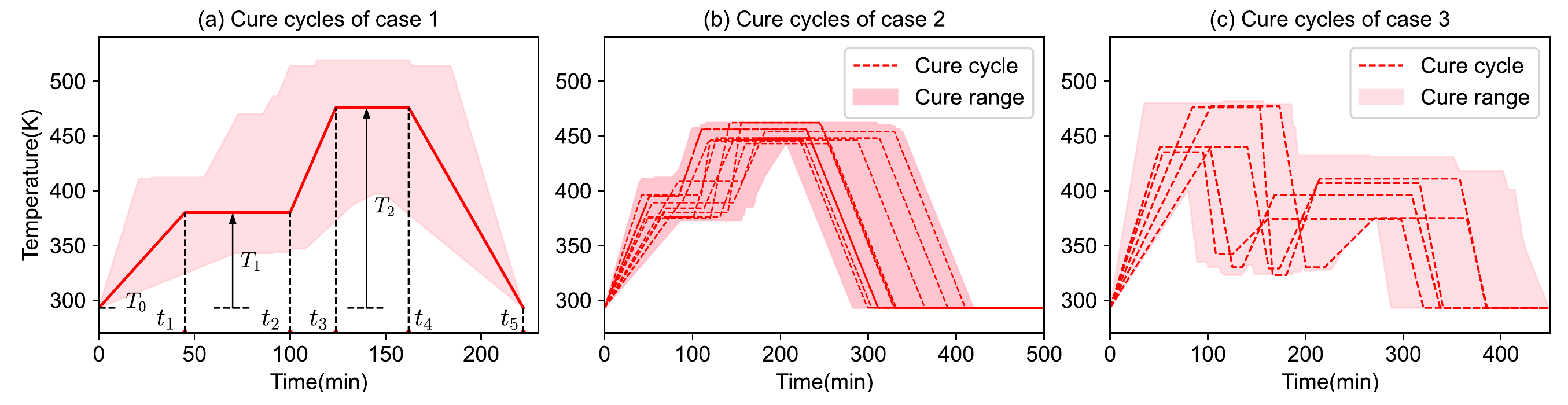}\\[-2mm]	
	\caption{Cure cycles for 3 cases}
	\label{fig:fig6}
\end{figure}

\paragraph{Training settings} The proposed framework is implemented in Pytorch, which is a famous open-source machine learning library. The structure and parameters of the network are designed based on the experiences of Kovachki et al. \cite{kovachki2021neural}, including 64 channels, 4 Fourier layers and 3 fullconnected output layers. The sizes of frequency modes in Fourier layer and residual mapping are both 16. Other detailed hyperparameters can be found in the submitted source code. The loss function is defined as the relevant L2 norm between predicted the temperature history $T_{pre}(t)_{i}$ and the real temperature history $T_{c}(t)_{i}$ :

\begin{equation}
{loss}_{L 2}=\frac{1}{n} \sum_{i=1}^{n} \frac{\left\|T_{c}(t)_{i}-T_{p re}(t)_{i}\right\|_{2}}{\left\|T_{c}(t)\right\|_{2}}
\end{equation}

All the experiments were carried out on an ordinary NVIDIA GeForce GTX 1660 SUPER GPU with 300 Epochs, a batch-size of 10 , and a learning rate of $0.001$. It is surprising that the training process only takes about $20 \mathrm{~s}$ and the prediction time can be negligible, thus the proposed method can be potentially used in online prediction during curing.

\subsection{Experimental results of case 1} 
\begin{figure}[th]
	\centering
	\includegraphics[width=1\linewidth]{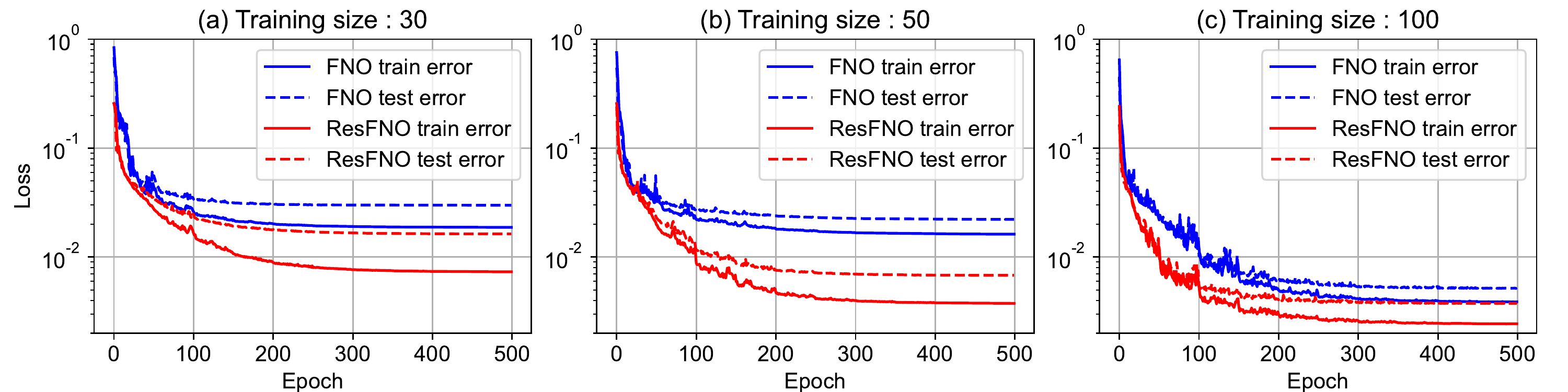}\\[-2mm]	
	\caption{The convergent results for FNO and ResFNO with different sizes of training data}
	\label{fig:fig7}
\end{figure}

In this case, 200 extreme cure cycles are generated randomly, and the corresponding temperature histories of the tool and composite part are simulated in COMSOL. The thickness of the composite-tool system (50mm) is discretized into 51 elements and the total cure time (222min) is discretized into 223 elements. Firstly, we build the prediction model of the temperature history of the mid-point of the composite part, namely $x=35 \mathrm{~mm}$. To check the influence of size of training data, $30,50,100$ groups of samples $\left[T_{a}(t), T_{c}(t)_{x=35 m m}\right]$ are randomly selected as training data to train FNO and ResFNO. The rest data are adopted as test data to study the generalization ability of the trained models. The convergences of train error and test error of FNO and ResFNO with different sizes of training data are shown Fig. \ref{fig:fig7}. It can be observed that ResFNO can converge quickly than FNO with different sizes of training data, and ResFNO can achieve much better performance with less training data. The relative L2 norm can be controlled below $0.01$ with only 50 data. Considering the marginal gain of loss shows a progressive decrease with the increasing of training data. We will select 50 as the size of training data for the subsequent experiments.

\begin{figure}[th]
	\centering
	\includegraphics[width=1\linewidth]{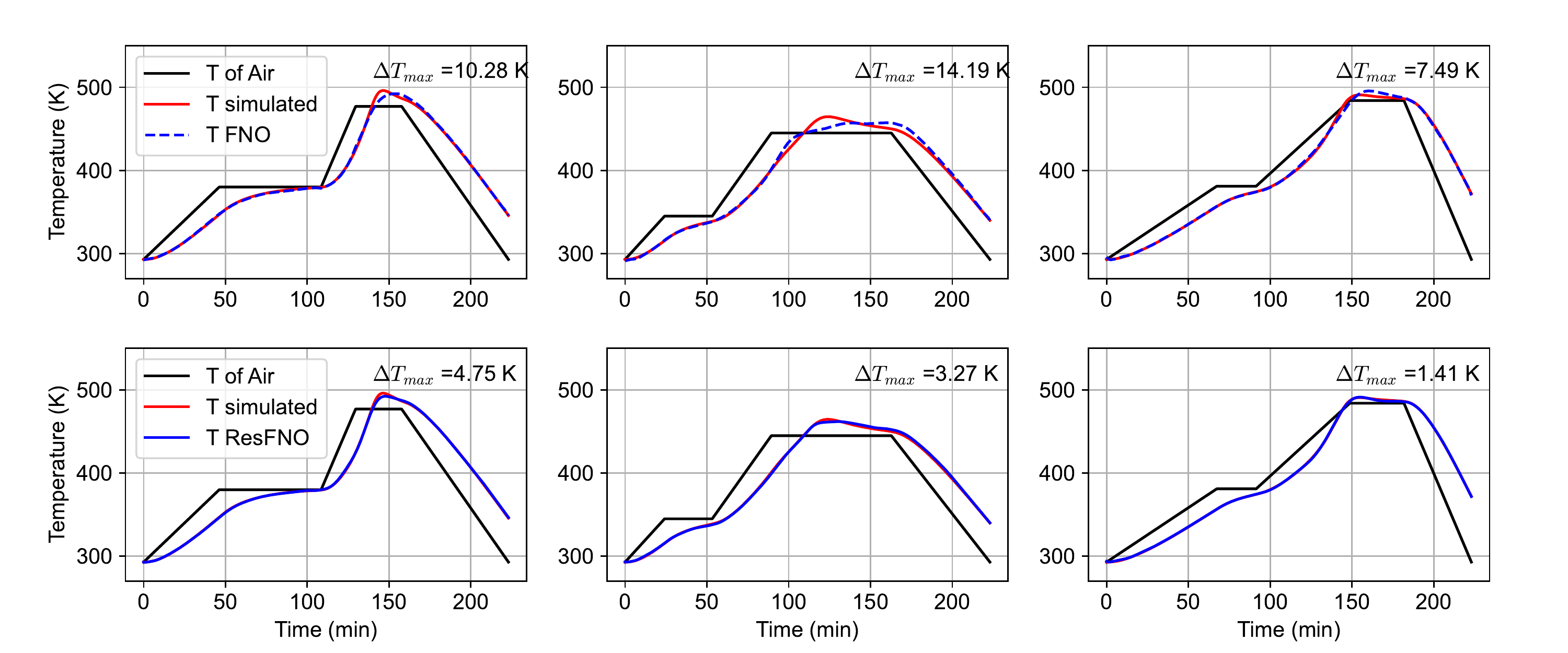}\\[-2mm]	
	\caption{The temperature histories predicted by FNO and ResFNO}
	\label{fig:fig8}
\end{figure}

\paragraph{Comparison between predicted temperature history} As the loss value cannot express the real fitting performance, 3 cure cycles in test datasets are randomly selected to test the prediction result explicitly. The designed cure cycle (T of Air), simulated temperature histories (T simulated) and predicted temperature histories (T FNO and T ResFNO) are depicted in Fig. \ref{fig:fig8}. It can be observed that both methods can provide satisfactory goodness of fit. To further study the performance of the proposed method, a quantitative index is necessary to defined according to the application requirements. As the prediction error of temperature may influence the degree of cure and mechanical property of resin, the maximum absolute prediction error $\Delta T_{\max }$ over the entire temperature history is selected to judge the prediction result. Compared with average prediction error shown in Fig. \ref{fig:fig7}, $\Delta T_{\max }$ is a strict criterion but more realistic for real application. As shown in Fig. \ref{fig:fig8}, $\Delta T_{\max }$ of FNO of 3 cure cycles are $10.28 K, 14.19 K$, and $7.49 \mathrm{~K}$ respectively. By contrast, ResFNO can reduce the $\Delta T_{\max }$ for all 3 cases to $4.75 \mathrm{~K}, 3.27 \mathrm{~K}$ and $1.41 \mathrm{~K}$.

\begin{figure}[th]
	\centering
	\includegraphics[width=1\linewidth]{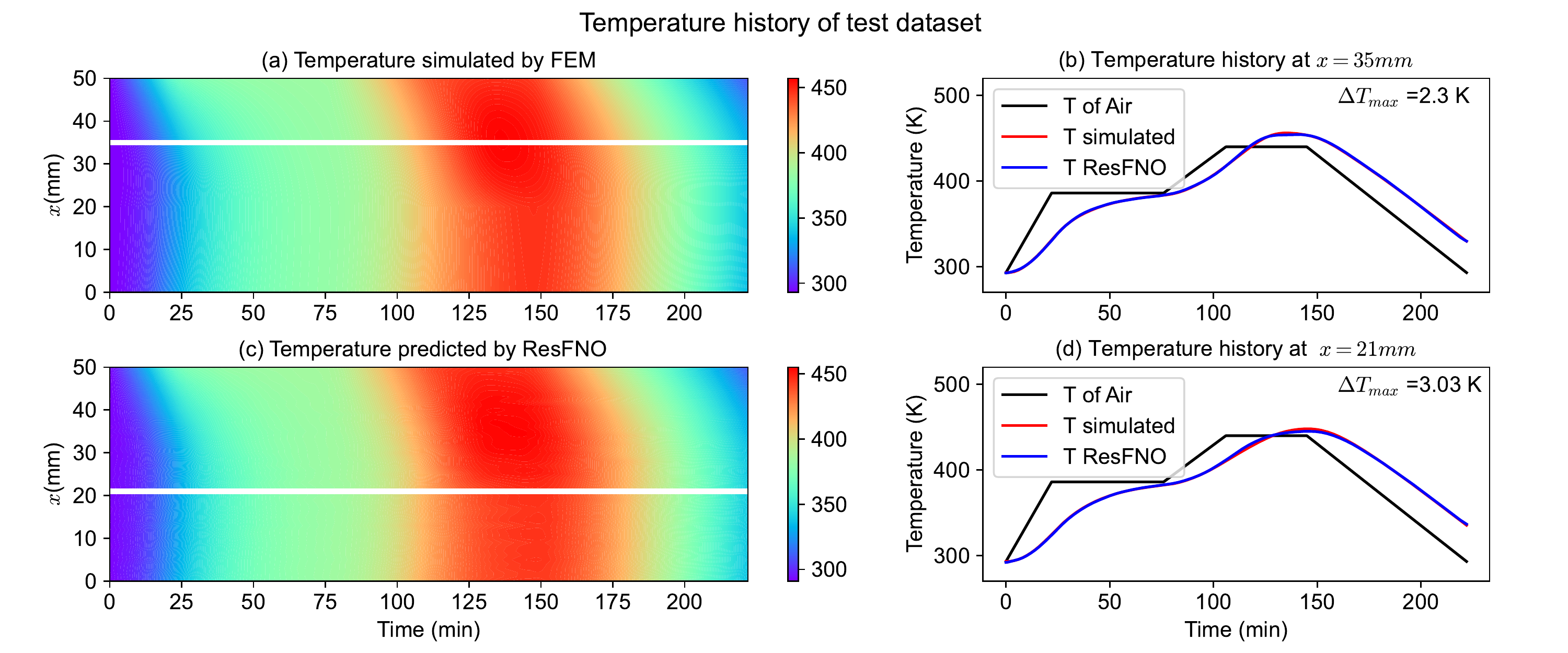}\\[-2mm]	
	\caption{Comparison between the predicted temperatures and the simulation results for case 1}
	\label{fig:fig9}
\end{figure}

\begin{figure}[th]
	\centering
	\includegraphics[width=1\linewidth]{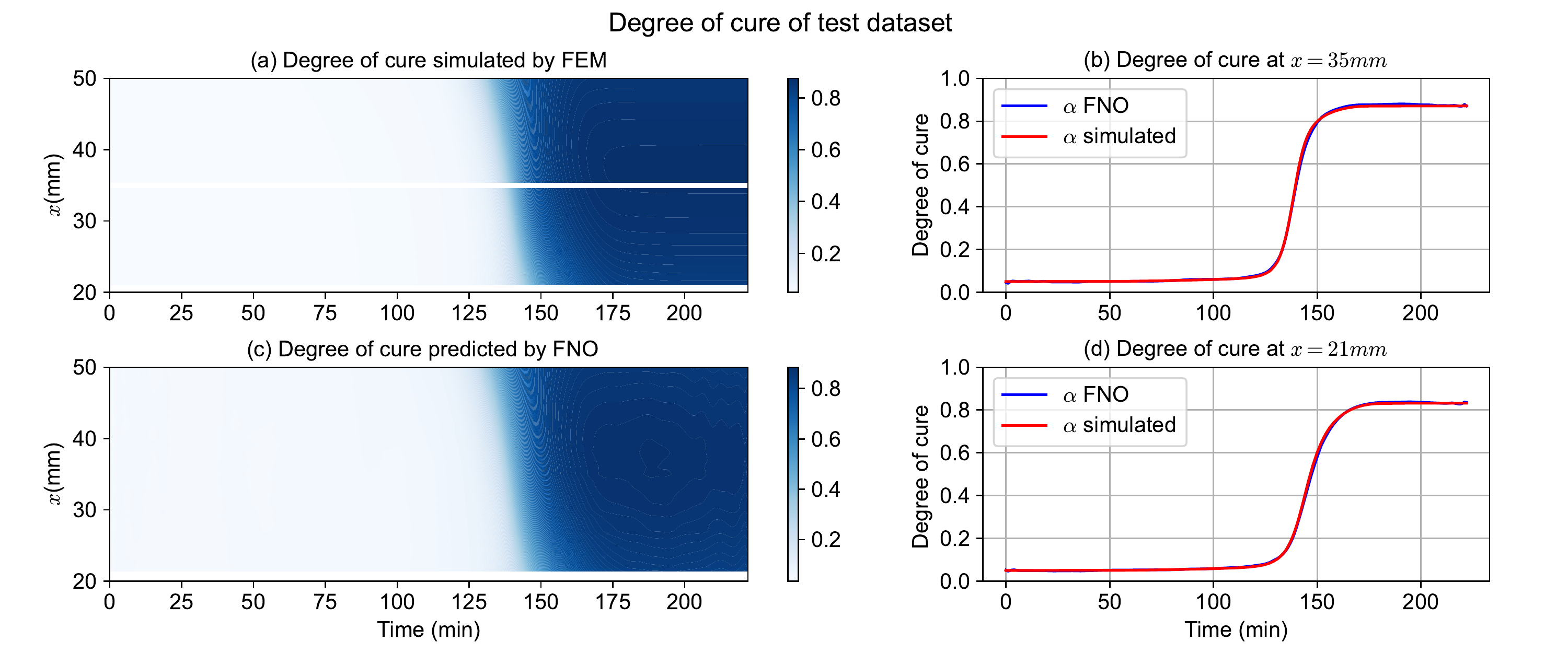}\\[-2mm]	
	\caption{Comparison between the predicted degree of cure and the simulation results for case 1}
	\label{fig:fig10}
\end{figure}

\paragraph{Predicted spatio-temporal temperature field } Abovementioned experiments result only focuses on $x=35 \mathrm{~mm}$, namely the mid-point of the composite part. To predict the temperature history of the whole composite-tool system, we have to train 51 sub-models from $x=0 \mathrm{~mm}$ to $x=51 \mathrm{~mm}$. The training effort is completely acceptable because the training process of one sub-model only takes about 20 s in a $\mathrm{PC}$ with GeForce GTX $1660 \mathrm{~S}$ GPU. The temperature histories of the whole composite-tool system predicted by FEM and ResFNO are shown in Fig. \ref{fig:fig9}(a) and Fig. \ref{fig:fig9}(c) respectively. The corresponding cure cycle is the black curve Fig. \ref{fig:fig9}(b) and Fig. \ref{fig:fig9}(d). And the predicted temperature histories for $x=35 \mathrm{~mm}$ and $x=21 \mathrm{~mm}$ (boundary of the composite part) are shown in Fig. \ref{fig:fig9}(b) and Fig. \ref{fig:fig9}(d) respectively. By the way, the predicted results of degree of cure are provided in Fig. \ref{fig:fig10}. Note that, these results are predicted by FNO rather than ResFNO because the proposed residual layer is not suitable for the prediction of degree of cure.

\begin{figure}[th]
	\centering
	\includegraphics[width=1\linewidth]{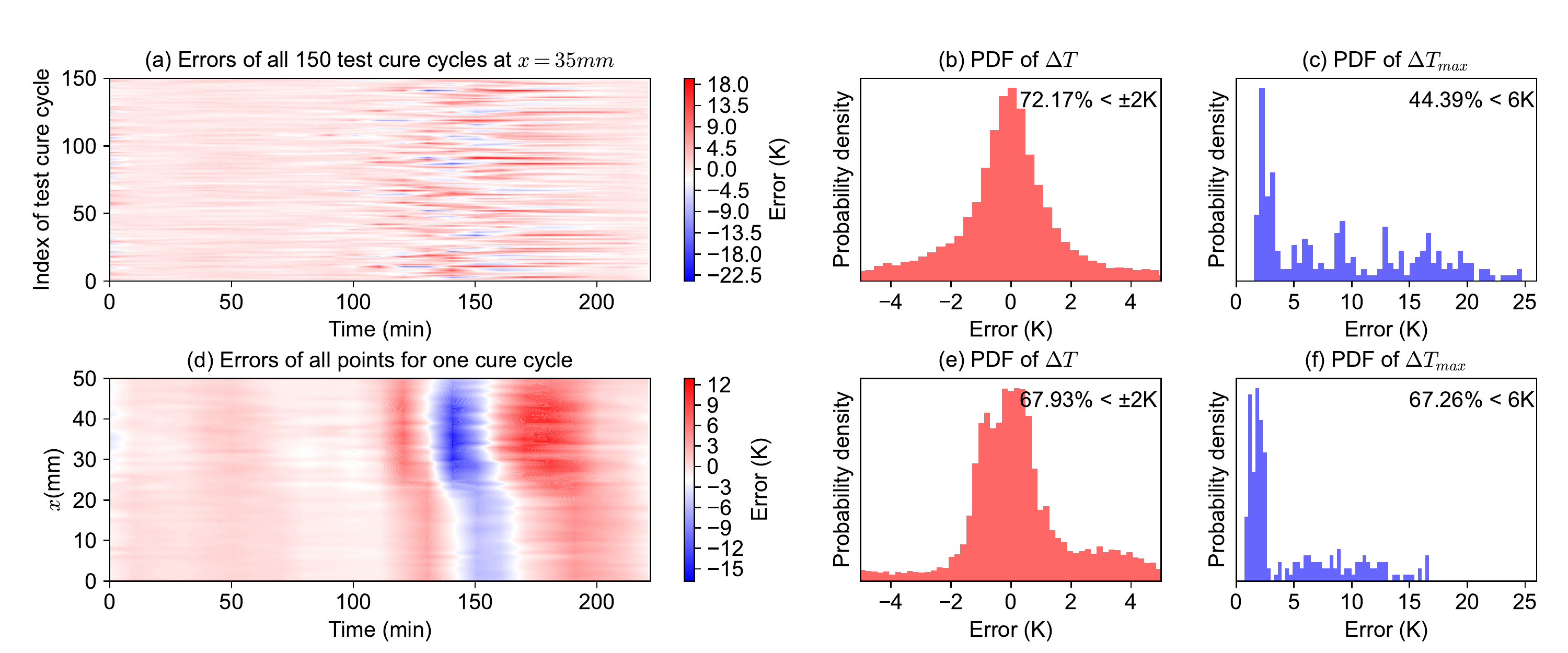}\\[-2mm]	
	\caption{Statistical analysis of prediction errors of FNO for case 1}
	\label{fig:fig11}
\end{figure}

\begin{figure}[th]
	\centering
	\includegraphics[width=1\linewidth]{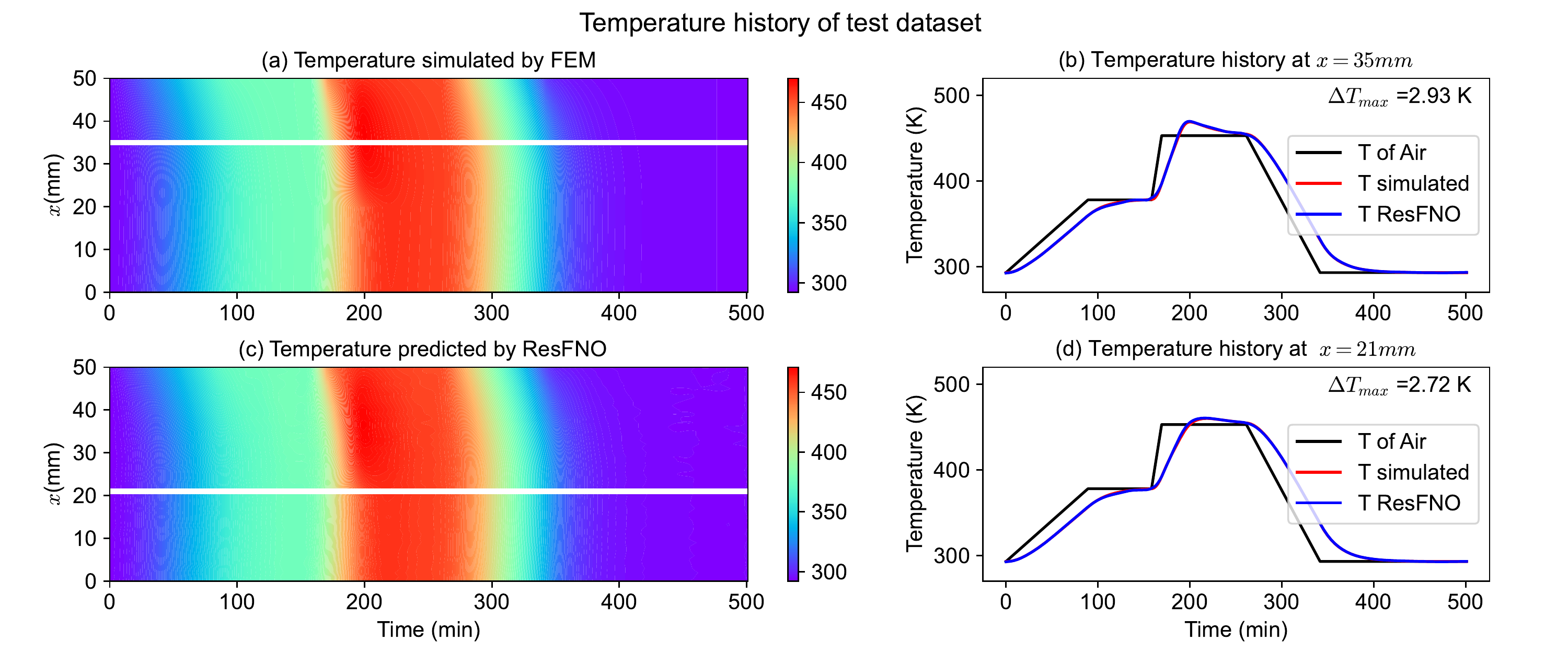}\\[-2mm]	
	\caption{Statistical analysis of prediction errors of ResFNO for case 1}
	\label{fig:fig12}
\end{figure}

\paragraph{Statistical analysis of prediction results } As Fig. \ref{fig:fig9} can only show the prediction results for one cure cycle, comprehensive statistical results are provided in Fig. \ref{fig:fig11} and Fig. \ref{fig:fig12} to further analyze the performance of the proposed ResFNO. The color map of prediction errors of all 150 test cure cycles at $x=35 \mathrm{~mm}$ is shown in Fig. \ref{fig:fig11}(a). And the colormap of prediction errors of $x=0 \mathrm{~mm}$ to $x=51 \mathrm{~mm}$ for one cure cycle is shown in Fig. \ref{fig:fig11}(d). The probability density of $\Delta T$ and $\Delta T_{\max }$ of the two colormaps are shown in Fig. \ref{fig:fig11}(b, c, e, f). It can be observed in Fig. \ref{fig:fig11} that the range of prediction error of FNO is $-22.5 \mathrm{~K}$ to $+18 \mathrm{~K}$. Only $44.39 \%$ cure cycles can obtain satisfactory prediction results with $\Delta T_{\max }<6 K$. By contrast, the prediction errors of ResFNO are much less than FNO as shown in Fig. \ref{fig:fig12}. The range of prediction error for all 150 test cure cycles are reduced to $-8 K$ to $+6.4 K$. Besides, the prediction errors of $96.19 \%$ points in Fig. \ref{fig:fig12}(a) are less than $\pm 2 K$, and $82.06 \%$ cure cycles can be predicted with $\Delta T_{\max }<6 K$. More indicators about prediction error can be found in Fig. \ref{fig:fig11} and Fig. \ref{fig:fig12}. It is clear that ResFNO can provide more accurate and stable prediction results compared with FNO.

\subsection{Experimental results of case 2}
\begin{figure}[th]
	\centering
	\includegraphics[width=1\linewidth]{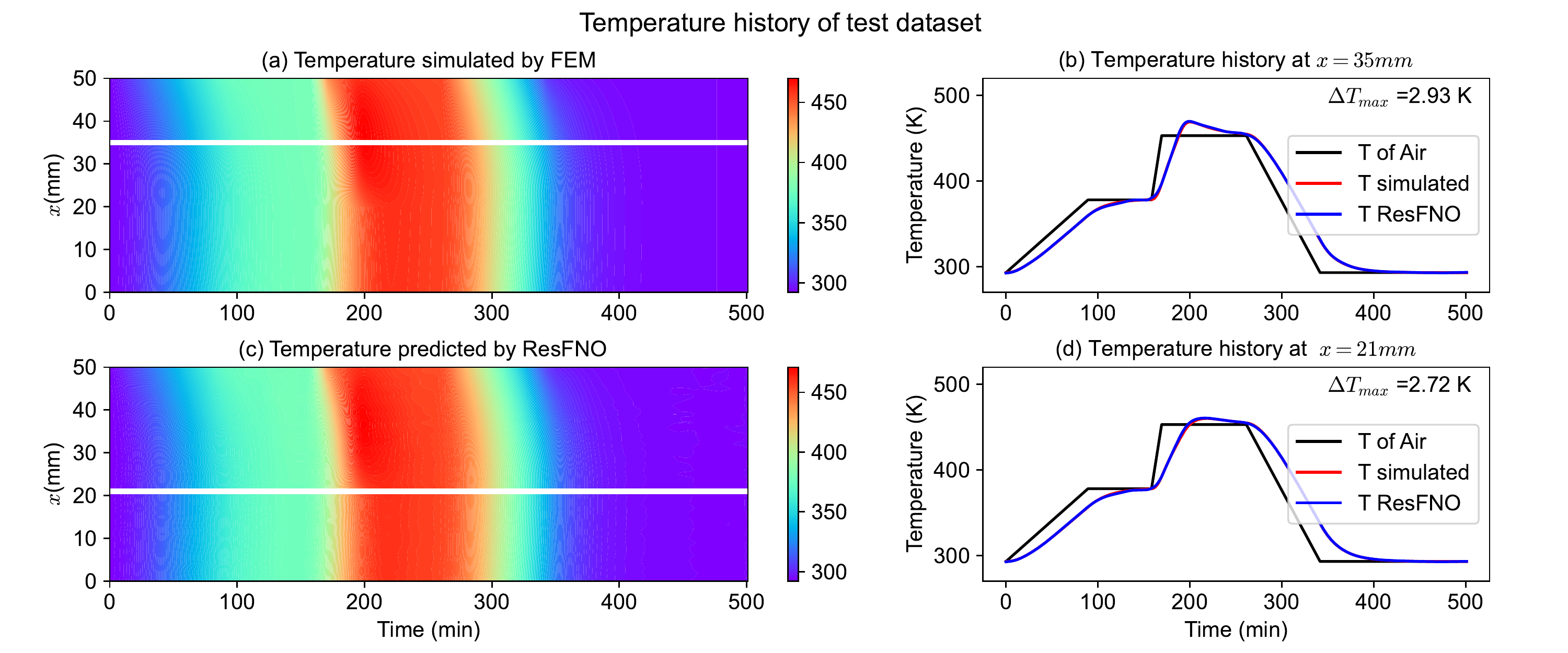}\\[-2mm]	
	\caption{Comparison between the predicted temperatures and the simulation results for case 2}
	\label{fig:fig13}
\end{figure}

\begin{figure}[th]
	\centering
	\includegraphics[width=1\linewidth]{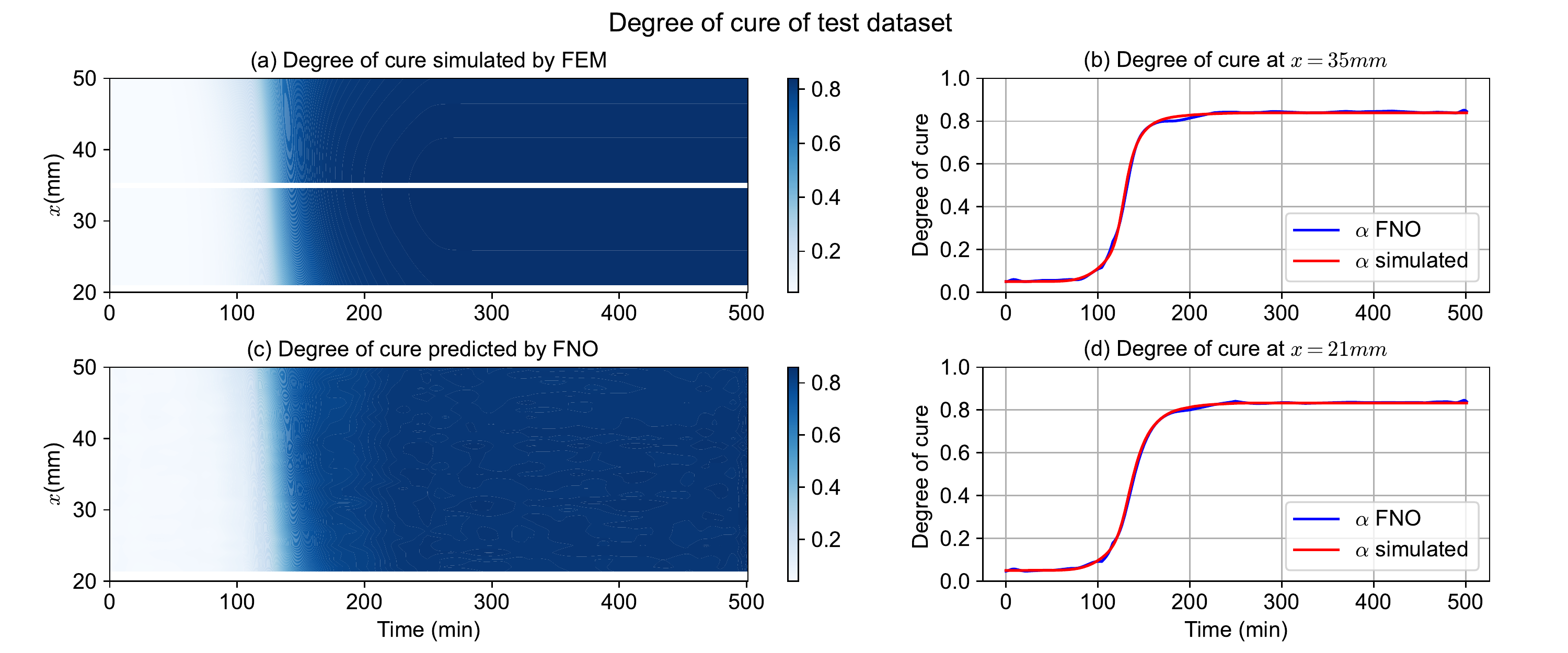}\\[-2mm]	
	\caption{Comparison between the predicted degree of cure and the simulation results for case 2}
	\label{fig:fig14}
\end{figure}

In this case, we will analyze the performance of ResFNO with more realistic cure cycles. These cure cycles are generated considering many physical constraints such as heating rate range of autoclave, post-cure temperature, holding time et. al. Most of experimental configures are the same with case1 including 50 training samples, 150 test samples and 51 discretized elements in the thickness direction of the composite-tool system. One difference is the variant curing time from around $270 \min$ to $420 \min$. Additional indoor temperatures $(293 \mathrm{~K})$ are added to the time series of cure cycles to ensure the identical size of $T_{a}(t)$, namely $500 \mathrm{~min}$ discretized into 501 elements. The temperature histories of the composite-tool system predicted by FEM and ResFNO are shown in Fig. \ref{fig:fig13}(a) and Fig. \ref{fig:fig13}(c) respectively. And the predicted temperature histories for $x=35 \mathrm{~mm}$ and $x=21 \mathrm{~mm}$ are shown in Fig. \ref{fig:fig13}(b) and Fig. \ref{fig:fig13}(d) respectively. The predicted results of ResFNO are awesome and the maximum prediction errors are only $2.93 K$ and $2.72 K$. The predicted results of degree of cure are provided in Fig. \ref{fig:fig14}. The maximum prediction errors for $x=21 \mathrm{~mm}$ and $x=51 \mathrm{~mm}$ are within the acceptable range $0.02$.

\begin{figure}[th]
	\centering
	\includegraphics[width=1\linewidth]{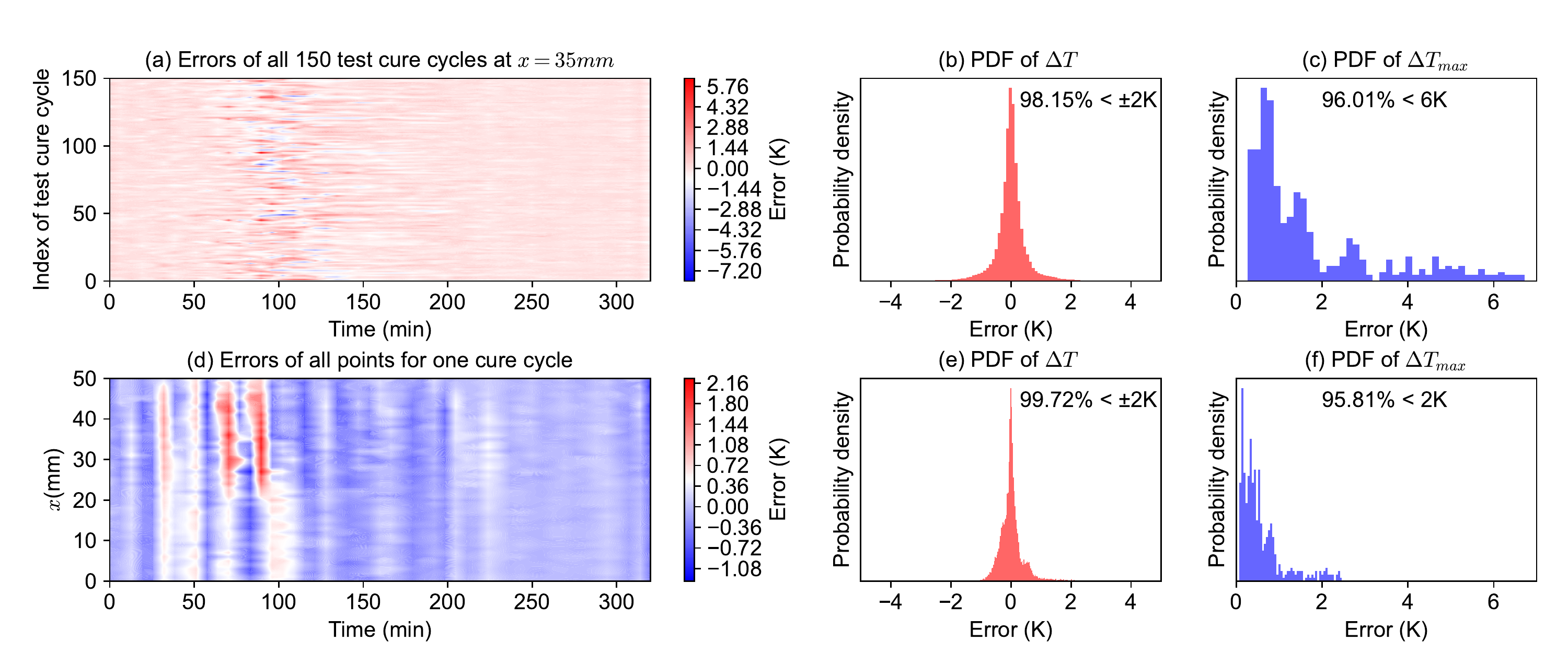}\\[-2mm]	
	\caption{Statistical analysis of prediction errors of ResFNO for case 2}
	\label{fig:fig15}
\end{figure}

The statistical analysis of prediction error of ResFNO are shown in Fig. \ref{fig:fig15}. The prediction errors of more than $98 \%$ points in all 150 test cycles at $x=35 \mathrm{~mm}$ are less than $\pm 2 K$. And almost all points from $x=0 \mathrm{~mm}$ to $x=51 \mathrm{~mm}$ for the selected cure cycle show errors less than $\pm 2 K$. More than $96 \% \Delta T_{\max }$ are less than $6 K$ besides few special cure cycles.

\subsection{Experimental results of case 3: transfer learning}
\begin{figure}[th]
	\centering
	\includegraphics[width=1\linewidth]{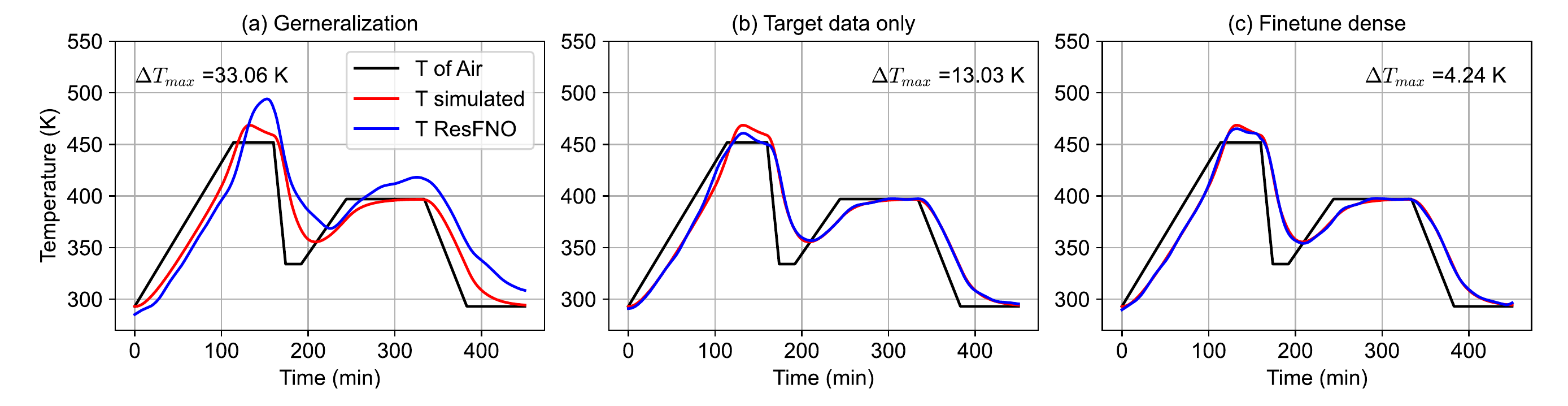}\\[-2mm]	
	\caption{Experimental results of the generalizability and transferability of ResFNO}
	\label{fig:fig16}
\end{figure}

Case 1 and Case 2 have demonstrated that ResFNO can provide satisfactory prediction results for classical two-hold cure cycles. In this case, we will explore the generalizability and trasferbility of the trained ResFNO in special smart cure cycles. Firstly, we train a ResFNO model using 50 samples in case1, then predict the temperature of a smart cure cycle to evaluate if the trained model can be generalized to more broader distribution. As shown in Fig. \ref{fig:fig16}(a), the predicted result is extremely terrible and $\Delta T_{\max }$ reaches $33.06 K$. The result mean that ResFNO model trained on 50 two-hold samples can only be used to prediction cure cycles from similar distribution.

Traditiaonal machine learning method can maintain its effectiveness only when all training data and test data follow the same distibution. When the distribution of target domain is different from the source domain, transfer learning methods can adapt the distribution discrepency with few labelled target data \cite{chen2020transfer}.Therefore, 10 smart cure cycles and the corresponding temperature histories are selected as target data to finetue the ResFNO network trained in case1. Considering the spectial structure of ResFNO, we designed three finetuing strategies, including finetuing dense layers (Lifting and Projection layers), finetuing Fourier Layers, and finetuing all parameters.

As shown in Fig. \ref{fig:fig16}(c), $\Delta T_{\max }$ of the predicted temperature history after finetuing dense layers is reduced to only $4.24 K$. The statistical anlysis results of 40 complex smart cure cycles are shown in Table. 3. It can be observed that transfer learning can improve the prediction results to some extent and finetuing dense layers seems to be the best strategy. The average $\Delta T_{\max }$ of 40 cycles after finetuing dense layers is around $8.64 \mathrm{~K}$. To be honest, the prediction results are not as satisfactory as case 1 and case 2 because of the extreme complex cure cycles and few available data. But these experiments show the potential of transfer learning for ResFNO, which will be our next concerned topic.

\begin{table}
	\caption{Statistical analysis of $\Delta T_{\max }$ on different transfer learning strategies}
	\centering
	\vspace{-0.5em}
    \begin{tabular}{cc}
     \toprule
      Strategy & Mean and Standard deviation of $\Delta T_{\max }(\mathrm{K})$ \\
    \midrule Generalization & $35.96 (6.25)$ \\
    Target data only & $13.39 (4.79)$ \\
    Finetune Fourier layers & $9.54 (3.31)$ \\
    Finetune dense layers & $8.64 (2.61)$ \\
    Finetune all parameters & $9.68 (3.37)$ \\
    \bottomrule
    \end{tabular}
	\label{tab:table3}
\end{table}

\section{Conclusion}

During the curing process of composites, the temperature history directly determines the evolutions of the field of degree of cure as well as the residual stress, which will further influence the mechanical properties of composite, thus it is important to simulate the real temperature history to optimize the curing process of composites. A Residual Fourier Neural Operator framework is proposed to establish a direct mapping from the cure cycle $T_{a}(t)$ to the temperature history $T_{c}(t)$. Several case studies have evaluated the significant performance of the proposed method. Some conclusions can be drawn as follows:
\begin{itemize}[leftmargin=*]
\item It is effective to model the mapping between the cure cycle to the temperature history using neural operators because the two temperature functions can be decomposed into limited number of modes in frequency domain.
\item  By incorporating the domain knowledge into neural operators, the proposed ResFNO can provide more accurate prediction results, accelerate the training process as well as reduce the requirements of training data.
\item  According to the experimental results of case 3 , the proposed ResFNO has shown the potential to transfer the learned knowledge to more complex scenario.
\end{itemize}
It will be of further interest to apply ResFNO to more complex settings including changing heat transfer coefficient and spatial-temporal field prediction, with the aim of developing more effective and general thermal modelling methodologies.

\bibliographystyle{unsrt}
\bibliography{main}

\begin{thebibliography}{10}

\bibitem{harris2002design}
Charles~E Harris, James~H Starnes~Jr, and Mark~J Shuart.
\newblock Design and manufacturing of aerospace composite structures,
  state-of-the-art assessment.
\newblock {\em Journal of aircraft}, 39(4):545--560, 2002.

\bibitem{crawford2021mini}
Bryn Crawford, Hamid Khayyam, Abbas~S Milani, et~al.
\newblock A mini-review and perspective on current best practice and emerging
  industry 4.0 methods for risk reduction in advanced composites manufacturing.
\newblock {\em Open Journal of Composite Materials}, 11(02):31, 2021.

\bibitem{chen2021effect}
Cheng Chen.
\newblock {\em The effect of cure cycle on microstructure and mechanical
  properties of interlayer toughened composites}.
\newblock PhD thesis, University of British Columbia, 2021.

\bibitem{struzziero2019numerical}
Giacomo Struzziero, Julie~JE Teuwen, and Alexandros~A Skordos.
\newblock Numerical optimisation of thermoset composites manufacturing
  processes: A review.
\newblock {\em Composites Part A: Applied Science and Manufacturing},
  124:105499, 2019.

\bibitem{hubert2012autoclave}
Pascal Hubert, G~Fernlund, and A~Poursartip.
\newblock Autoclave processing for composites.
\newblock In {\em Manufacturing techniques for polymer matrix composites
  (PMCs)}, pages 414--434. Elsevier, 2012.

\bibitem{carrera2002theories}
E19178181062 Carrera.
\newblock Theories and finite elements for multilayered, anisotropic, composite
  plates and shells.
\newblock {\em Archives of Computational Methods in Engineering}, 9(2):87--140,
  2002.

\bibitem{fernlund2003finite}
G~Fernlund, A~Osooly, A~Poursartip, R~Vaziri, R~Courdji, K~Nelson, P~George,
  L~Hendrickson, and J~Griffith.
\newblock Finite element based prediction of process-induced deformation of
  autoclaved composite structures using 2d process analysis and 3d structural
  analysis.
\newblock {\em Composite Structures}, 62(2):223--234, 2003.

\bibitem{zhang2021physical}
Meng Zhang, Fei Tao, Biqing Huang, and AYC Nee.
\newblock A physical model and data-driven hybrid prediction method towards
  quality assurance for composite components.
\newblock {\em CIRP Annals}, 2021.

\bibitem{chen2019pose}
Gengxiang Chen, Yingguang Li, and Xu~Liu.
\newblock Pose-dependent tool tip dynamics prediction using transfer learning.
\newblock {\em International Journal of Machine Tools and Manufacture},
  137:30--41, 2019.

\bibitem{humfeld2021machine}
Keith~D Humfeld, Dawei Gu, Geoffrey~A Butler, Karl Nelson, and Navid Zobeiry.
\newblock A machine learning framework for real-time inverse modeling and
  multi-objective process optimization of composites for active manufacturing
  control.
\newblock {\em arXiv preprint arXiv:2104.11342}, 2021.

\bibitem{zobeiry2021physics}
Navid Zobeiry and Keith~D Humfeld.
\newblock A physics-informed machine learning approach for solving heat
  transfer equation in advanced manufacturing and engineering applications.
\newblock {\em Engineering Applications of Artificial Intelligence},
  101:104232, 2021.

\bibitem{ramezankhani2021making}
Milad Ramezankhani, Bryn Crawford, Apurva Narayan, Heinz Voggenreiter, Rudolf
  Seethaler, and Abbas~S Milani.
\newblock Making costly manufacturing smart with transfer learning under
  limited data: A case study on composites autoclave processing.
\newblock {\em Journal of Manufacturing Systems}, 59:345--354, 2021.

\bibitem{ramezankhani2021active}
Milad Ramezankhani, Apurva Narayan, Rudolf Seethaler, and Abbas~S Milani.
\newblock An active transfer learning (atl) framework for smart manufacturing
  with limited data: Case study on material transfer in composites processing.
\newblock In {\em 2021 4th IEEE International Conference on Industrial
  Cyber-Physical Systems (ICPS)}, pages 277--282. IEEE, 2021.

\bibitem{zobeiry2021theory}
Navid Zobeiry and Anoush Poursartip.
\newblock Theory-guided machine learning for process simulation of advanced
  composites.
\newblock {\em arXiv preprint arXiv:2103.16010}, 2021.

\bibitem{chen2021physics}
Gengxiang Chen, Yingguang Li, Xu~Liu, and Bo~Yang.
\newblock Physics-informed bayesian inference for milling stability analysis.
\newblock {\em International Journal of Machine Tools and Manufacture},
  167:103767, 2021.

\bibitem{raissi2017physics}
Maziar Raissi, Paris Perdikaris, and George~Em Karniadakis.
\newblock Physics informed deep learning (part i): Data-driven solutions of
  nonlinear partial differential equations.
\newblock {\em arXiv preprint arXiv:1711.10561}, 2017.

\bibitem{niaki2021physics}
Sina~Amini Niaki, Ehsan Haghighat, Trevor Campbell, Anoush Poursartip, and Reza
  Vaziri.
\newblock Physics-informed neural network for modelling the thermochemical
  curing process of composite-tool systems during manufacture.
\newblock {\em Computer Methods in Applied Mechanics and Engineering},
  384:113959, 2021.

\bibitem{kovachki2021neural}
Nikola Kovachki, Zongyi Li, Burigede Liu, Kamyar Azizzadenesheli, Kaushik
  Bhattacharya, Andrew Stuart, and Anima Anandkumar.
\newblock Neural operator: Learning maps between function spaces.
\newblock {\em arXiv preprint arXiv:2108.08481}, 2021.

\bibitem{he2016deep}
Kaiming He, Xiangyu Zhang, Shaoqing Ren, and Jian Sun.
\newblock Deep residual learning for image recognition.
\newblock In {\em Proceedings of the IEEE conference on computer vision and
  pattern recognition}, pages 770--778, 2016.

\bibitem{springer1967thermal}
George~S Springer and Stephen~W Tsai.
\newblock Thermal conductivities of unidirectional materials.
\newblock {\em Journal of Composite Materials}, 1(2):166--173, 1967.

\bibitem{soohyun2015smart}
Nam Soohyun, Lee Dongyoung, Choi Ilbeom, and Lee Dai~Gil.
\newblock Smart cure cycle for reducing the thermal residual stress of a
  co-cured e-glass/carbon/epoxy composite structure for a vanadium redox flow
  battery.
\newblock {\em Composite Structures}, 120:107--116, 2015.

\bibitem{chen2020transfer}
Gengxiang Chen, Yingguang Li, and Xu~Liu.
\newblock Transfer learning under conditional shift based on fuzzy residual.
\newblock {\em IEEE Transactions on Cybernetics}, 2020.

\end{thebibliography}



\end{document}